\documentclass[aps,twocolumn,prb,showpacs,superscriptaddress]{revtex4-1}
\usepackage{makeidx}
\usepackage{amsfonts}
\usepackage{graphics}
\usepackage{graphicx}
\usepackage{amssymb}
\usepackage{amsthm}
\usepackage{amsmath}
\usepackage{hyperref}

\begin{document}

\title{Phase diagram of a model for topological superconducting wires}
\author{D. P\'erez Daroca}
\affiliation{Gerencia de Investigaci\'on y Aplicaciones, Comisi\'on Nacional de
Energ\'ia At\'omica, 1650 San Mart\'{\i}n, Buenos Aires, Argentina} 
\affiliation{Consejo Nacional de Investigaciones Cient\'{\i}ficas y T\'ecnicas,
1025 CABA, Argentina}

\author{A. A. Aligia}
\affiliation{Centro At\'{o}mico Bariloche, Comisi\'{o}n Nacional
de Energ\'{\i}a At\'{o}mica, 8400 Bariloche, Argentina} 
\affiliation{Instituto Balseiro, Comisi\'{o}n Nacional
de Energ\'{\i}a At\'{o}mica, 8400 Bariloche, Argentina} 
\affiliation{Consejo Nacional de Investigaciones Cient\'{\i}ficas y T\'ecnicas,
1025 CABA, Argentina}

\begin{abstract}
We calculate the phase diagram of a model for topological superconducting
wires with local s-wave pairing, spin-orbit coupling $\vec{\lambda}$ and
magnetic field $\vec{B}$ with arbitrary orientations. This model is a
generalized lattice version of the one proposed by Lutchyn \textit{et al.}
[Phys. Rev. Lett. \textbf{105} 077001 (2010)] and Oreg \textit{et al.}
[Phys. Rev. Lett. \textbf{105} 177002 (2010)], who considered $\vec{\lambda}$
perpendicular to $\vec{B}$. The model has a topological gapped phase with
Majorana zero modes localized at the ends of the wires. We determine
analytically the boundary of this phase. When the directions of the
spin-orbit coupling and magnetic field are not perpendicular, in addition to the
topological phase and the gapped nontopological phase, a  
gapless superconducting phase appears.
\end{abstract}

\maketitle


\section{Introduction}

\label{intro}

The study of topological superconducting wires, which host Majorana zero
modes (MZMs) at their ends, is a field of intense research in condensed
matter physics, not only because of the interesting basic physics involved 
\cite{sato}, but also because of possible applications in decoherence-free
quantum computing.\cite{kitaev,nayak,alicea,lobos}

In 2010, Lutchyn \textit{et al.} \cite{lutchyn2010} and Oreg \textit{et al.} 
\cite{oreg2010} proposed a model for topological superconducting wires
describing a system formed by a semiconducting wire with spin-orbit coupling
(SOC) and proximity-induced s-wave superconductivity under an applied
magnetic field perpendicular to the direction of the SOC. This yields a
topological superconducting phase with MZMs localized at its ends. The
observation of these MZMs in these types of wires was reported in different
experimental studies.\cite{wires-exp1,wires-exp2,wires-exp3,wires-exp4}

The search for different models and mechanisms leading to topological
superconducting phases continues being a very active avenue of research
theoretically and experimentally.

More recently, there has been experimental research as well as theoretical
studies in similar models, including those for time-reversal invariant
topological superconductors,\cite{review-tritops,volpez} of the effects of
MZMs in Josephson junctions, in particular because the dependence on the
applied magnetic flux introduces an additional control knob.\cite{volpez,zazu,pientka,hell,ren,fornie,cata,tomo}

In particular, it has been recently proposed that the current-phase relation
measured in Josephson junctions may be used to find the parameters that
define the MZMs.\cite{tomo} A possible difficulty in these experiments is
the slow thermalization to the ground state
in the presence of a gap.\cite{bondy}
A way to circumvent this problem is to rotate the magnetic field
slowly from a direction not perpendicular to the SOC in which the system is
in a gapless superconducting phase, in which thermalization is easier.\cite{tomo}
Therefore, it is convenient to know the phase diagram of the system
and the extension of this gapless phase.

In this work we calculate the phase diagram of the lattice version of the
model and discuss in particular the gapless phase. The paper is organized as
follows. In Sec. \ref{model} we describe the model. The topological
invariants use to define the phase diagram are presented in Sec. \ref{inv}. In Sec. \ref{res} we show the numerical results, analytical expressions for
the boundaries of the topological phase and discuss briefly the Majorana zero modes.
We summarize the results in Sec. \ref{sum}.

\section{Model}

\label{model}

The model for topological superconducting wires studied in this work is the
lattice version of that introduced by Lutchyn \textit{et al.} 
\cite{lutchyn2010} and Oreg \textit{et al.} \cite{oreg2010}. The Hamiltonian can
be written as \cite{tomo} 
\begin{eqnarray}
H &=&\sum_{\ell }[\mathbf{c}_{\ell }^{\dagger }\left( -t\;\sigma _{0}-i\vec{
\lambda}\cdot \vec{\sigma}\right) \mathbf{c}_{\ell +1}+\Delta c_{\ell
\uparrow }^{\dagger }c_{\ell \downarrow }^{\dagger }+\text{H.c.}  \notag \\
&-&\mathbf{c}_{\ell }^{\dagger }\left( \vec{B}\cdot \vec{\sigma}+\mu \sigma
_{0}\right) \mathbf{c}_{\ell }],\;\;  \label{ham}
\end{eqnarray}
where $\ell $ labels the sites of a chain, $\mathbf{c}_{\ell }=(c_{\ell
\uparrow },c_{\ell \downarrow })^{T}$, $t$ is the nearest-neighbor hopping, $
\vec{\lambda}$ is the SOC, $\Delta $ represents the magnitude of the
proximity-induced superconductivity, $\vec{B}$ is the applied magnetic field
and $\mu $ is the chemical potential. As usual, the components of the vector $
\vec{\sigma}=\left( \sigma _{x},\sigma _{y},\sigma _{z}\right) $ are the
Pauli matrices and $\sigma _{0}$ is the 2$\times $2 unitary matrix. The
pairing amplitude $\Delta $ can be assumed real. Otherwise, the phase can be
eliminated by a gauge transformation in the operators $c_{\ell \sigma
}^{\dagger }$ that absorbs the phase.

Without loss of generality, we choose the $z$ direction as that of the
magnetic field ($\vec{B}=B\mathbf{\hat{z}}$) and $x$ perpendicular to the
plane defined by $\vec{\lambda}$ and $\vec{B}$ ($\vec{\lambda}=\lambda _{y} 
\mathbf{\hat{y}}+\lambda_{z} \mathbf{\hat{z}}$). After Fourier
transformation, the Hamiltonian takes the form $H=\sum_{k}H_{k}$, with

\begin{eqnarray}
H_{k} &=&-(\mu +2t\cos (k))(c_{k\uparrow }^{\dagger }c_{k\uparrow
}+c_{k\downarrow }^{\dagger }c_{k\downarrow }) \\
&& -B(c_{k\uparrow }^{\dagger }c_{k\uparrow }-c_{k\downarrow }^{\dagger
}c_{k\downarrow }) -2\sin (k)\left[ i\lambda _{y}(c_{k\uparrow }^{\dagger
}c_{k\downarrow }-c_{k\downarrow }^{\dagger }c_{k\uparrow }) \right.  \notag
\\
&& \left. +\lambda _{z}(c_{k\uparrow }^{\dagger }c_{k\uparrow
}-c_{k\downarrow }^{\dagger }c_{k\downarrow })\right] +\Delta (c_{k\uparrow
}^{\dagger }c_{-k\downarrow }^{\dagger }+c_{-k\downarrow }c_{k\uparrow }). 
\notag  \label{hk}
\end{eqnarray}
Using the four-component spinor $(c_{k\uparrow }^{\dagger },c_{k\downarrow
}^{\dagger },c_{-k\uparrow },c_{-k\downarrow })$,\cite{tewari} the
contribution to the Hamiltonian for wave vector $k$ can be written in the
form

\begin{eqnarray}
H_{k} &=&-(\mu +2t\cos (k))\tau _{z}\otimes \sigma _{0}-B\tau _{z}\otimes
\sigma _{z}-\Delta \tau _{y}\otimes \sigma _{y}  \notag \\
&&+2\lambda _{y}\sin (k)\tau _{z}\otimes \sigma _{y}-2\lambda _{z}\sin
(k)\tau _{0}\otimes \sigma _{z},  \label{hk2}
\end{eqnarray}%
where the Pauli matrices $\sigma _{\alpha }$ act on the spin space, while
the $\tau _{\alpha }$ act on the particle-hole space. Writing the matrix
explicitly, $H_{k}$ takes the form

\begin{equation}
H_{k}=%
\begin{pmatrix}
-a-B-z & -iy & 0 & \Delta \\ 
iy & -a+B+z & -\Delta & 0 \\ 
0 & -\Delta & a+B-z & iy \\ 
\Delta & 0 & -iy & a-B+z
\end{pmatrix}
\label{hk3}
\end{equation}%
where $a=\mu +2t\cos (k)$, $B=|B|=B_{z}$, $y=2\lambda _{y}\sin (k)$ and $
z=2\lambda _{z}\sin (k)$.

\section{Topological invariants}

\label{inv} In this section we define the topological invariants we use to
characterize the topological phases. In general, the Hamiltonian belongs to
topological class D with a 
$\mathbb{Z}_{2}$ topological invariant.\cite{Schn,ryu} However, for perpendicular $\vec{\lambda}$ and $\vec{B}$ ($z=0$),
the system has a chiral symmetry and belongs to the topological class BDI
with a $\mathbb{Z}$ (integer) topological invariant corresponding to a
winding number.\cite{tewari}  In this case, the calculation of the
topological invariant is simpler, as shown by Tewari and Sau.\cite{tewari}  

Following this work, we perform a rotation in $\pi /2$ around the $\hat{y}$
axis in particle-hole space, which transforms $\tau _{z}$ to $\tau _{x}$: $
H_{k}^{\prime }=UH_{k}U^{\dagger }$ with $U=$exp$(-i\pi /4)\tau _{y}$. With
this transformation $H_{k}^{\prime }$ becomes

\begin{equation}
H_{k}^{\prime }=
\begin{pmatrix}
-z & 0 & -a-B & \Delta -iy \\ 
0 & z & -\Delta +iy & -a+B \\ 
-a-B & -\Delta -iy & -z & 0 \\ 
\Delta +iy & -a+B & 0 & z
\end{pmatrix}
\label{hkp}
\end{equation}
Taking  $z=0$, this rotation yields an off-diagonal (chiral symmetric)
Hamiltonian. This allows us to define a winding number $W$ (a topological $
\mathbb{Z}$ invariant) from the phase of the determinant of the $2\times 2$
matrix $A(k)$, which is the upper right corner of Eq. (\ref{hkp}).\cite{tewari}
Specifically $\mathrm{Det}(A(k))$$=|\mathrm{Det}(A(k))|$$e^{i\theta (k)}=$$a^{2}-B^{2}-(\Delta -iy)^{2}$, and

\begin{equation}
W=\frac{-i}{\pi }\int\limits_{0}^{\pi }\frac{d(e^{i\theta (k)})}{e^{i\theta
(k)}}.  \label{win}
\end{equation}
In addition, a $\mathbb{Z}_{2}$ invariant $I$ can be defined from the
relative sign of $\mathrm{Det}(A)$ (which is real for $k=0$ and $k=\pi$)
between the points $k=0$ and $k=\pi$:

\begin{equation}
I=(-1)^{W}=\text{sign}\frac{\mathrm{Det}(A(\pi ))}{\mathrm{Det}(A(0))}
\label{z2}
\end{equation}
Looking for the condition that $I\equiv -1$ (mod 2), we obtain that the
conditions for the system to be in the topological phase are that $\lambda
_{y}=|\vec{\lambda}|\neq 0\neq \Delta $ and the remaining parameters should
satisfy 
\begin{equation}
|2|t|-r|<|\mu |<|2|t|+r|,\text{ \ \ \ with }r=\sqrt{B^{2}-\Delta ^{2}}>0.
\label{bound}
\end{equation}
We note that changing the sign of any of the parameters does not change the
boundary of the topological phase. This is due to the symmetry properties of
the Hamiltonian.\cite{tomo}

In the more general case, when $\vec{\lambda}$ and $\vec{B}$ are not
perpendicular, it is not possible to follow the approach outlined above. In
this case, we use the Zak Berry phase to construct the topological
invariant. \cite{zak,king,resta,ortiz,bf1,bf2,hatsu,bf3,ryu,deng,budich}
Specifically, the Hamiltonian $H_{k}$ has four different eigenvectors and
for each of them, following Zak,\cite{zak} one can calculate a Berry phase
from the Bloch functions as the wave vector $k$ varies in the loop $0\leq
k\leq 2\pi $ (with $k=2\pi $ equivalent to $k=0$). For each eigenstate $
|u(k)\rangle $ of $H_{k}$, the Berry phase is 

\begin{equation}
\gamma =-\text{Im }\int\limits_{0}^{2\pi }dk\langle u(k)|\frac{\partial }{
\partial k}|u(k)\rangle .  \label{gam}
\end{equation}

\ In addition (as noted before \cite{tomo}) choosing a suitable coordinate
frame ( $\vec{\lambda}\cdot \mathbf{\hat{y}}=\vec{B}\cdot \mathbf{\hat{y}}=0$
), the Hamiltonian Eq. (\ref{ham}) is invariant under an antiunitary
operator defined as the product of inversion (defined by the transformation $
\ell \leftrightarrow N+1-\ell $, for a chain with $N$ sites) and complex
conjugation, implying that the Berry phase $\gamma $ is quantized with only
two possible values $0$ and $\pi $ (mod $2\pi $).\cite{hatsu} Naturally the
value of the Berry phase does not depend on the choice of the reference
frame. Therefore, as for an insulator, if the system has a gap, the sum of
the Berry phases of all one-particle states of energies below the gap mod $2\pi $, defines a 
$\mathbb{Z}_{2}$ topological number, indicating that the system is trivial
(topological) if this sum is equivalent to 0 ($\pi $) mod $2\pi .
$\cite{ryu,budich} Moreover, from Eq. (\ref{hk}) it is easy to realize that the
charge conjugation $c_{\ell \sigma }^{\dagger }\leftrightarrow c_{\ell
\sigma }$, which in Fourier space means $c_{k\sigma }^{\dagger }=(1/\sqrt{N}%
)\sum_{l}e^{-ik\ell }c_{\ell \sigma }^{\dagger }\leftrightarrow c_{-k\sigma }
$, transforms $H_{k}\leftrightarrow -H_{-k}$. Therefore the sum of the Berry
phases of all positive eigenvalues gives the same topological number as the
sum of all negative eigenvalues.

In our model, $H_{k}$ has four eigenvalues $E(k)$. The lowest one $E_{1}(k)$
is always negative and the corresponding eigenvector has always a Berry
phase 0. From the above mentioned charge-transfer symmetry, the fourth
eigenvalue (the highest one) has energy $E_{4}(k)=-E_{1}(-k)>0$. Therefore,
the Berry phase of the second eigenvalue (which is equal to that of the
third one) determines the $\mathbb{Z}_{2}$ invariant. We have calculated the
Berry phase $\gamma $ of each of the four bands (and particularly the second
one) from the normalized eigenvectors $|u_{j}\rangle =|u(k_{j})\rangle $ of
the $4\times 4$ matrix obtained numerically at $M$ wave vectors $k_{j}=2\pi
(j-1)/M$, using a numerically invariant expression \cite{ortiz,deng}. This
expression is derived in the following way. Discretizing Eq. (\ref{gam}) and
approximating $\partial |u(k)/\partial k=(M/2\pi )(|u(k_{j+1})\rangle
-|u(k_{j})\rangle )$, one obtains

\begin{equation}
\gamma =-\text{Im}\sum_{j=1}^{M}\left[ \langle u_{j}|\left( |u_{j+1}\rangle
-|u_{j}\rangle \right) \right] .  \label{gam2}
\end{equation}
If $M$ is large enough so that $k_{j}$ and $k_{j+1}$ are very close, then 
$x=\langle u_{j}|u_{j+1}\rangle -1$ is very small and one can retain only the
first term in the Taylor series expansion ln$(1+x)=x-x^{2}/2+...$ Replacing in Eq. (
\ref{gam2}) one obtains

\begin{eqnarray}
\gamma  &=&-\mathrm{{Im}\left[ \ln (P)\right] }\text{, where }  \notag \\
P &=&\langle u_{1}|u_{2}\rangle \langle u_{2}|u_{3}\rangle ...\langle
u_{M-1}|u_{1}\rangle   \label{gam4}
\end{eqnarray}
It is easy to see that Eq. (\ref{gam}) is gauge invariant. This means that
the result does not change if $|u(k)\rangle $ is replaced by $e^{i\varphi
(k)}|u(k)\rangle $, where $\varphi (k)$ is a smooth function with $\varphi
(2\pi )=\varphi (0).$ Similarly, the product $P$ is independent of the base
chosen by the numerical algorithm to find the eigenstates $|u_{j}\rangle .$
Therefore Eq. (\ref{gam4}) is numerically gauge invariant. Analyzing the
change in the results with increasing $M$, we find that  $M\sim 250$ is
enough to obtain accurately all phase boundaries shown below. A further
increase in  $M$ leads to changes that are not visible in the scale of the
figures.

This $\mathbb{Z}_{2}$ topological invariant defined by the Berry phase 
of the second (or third) state can be trivially extended to the gapless case 
if the energies of the second and third state do not cross as a function of $k$.
Even if the energies cross the Berry phases can be calculated switching 
the states at the crossing. However, this case is not of interest here.

\section{Results}

\label{res}

\subsection{Phase diagram}

\label{phdi}

\begin{figure}[h!]
\begin{center}
\includegraphics[width=\columnwidth]{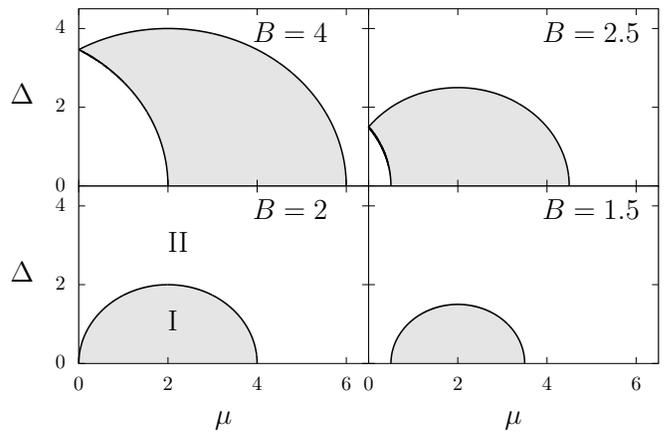}
\end{center}
\caption{Phase diagram in the $\mu,\Delta$ plane for perpendicular 
$\vec{\lambda}$ and $\vec{B}$, $t=1$, 
$\lambda=|\vec{\lambda}|=2$, and several values of $B$. 
Gray region I denotes the topological sector and white region II the 
non topological one.}
\label{variob}
\end{figure}

We start by discussing the simplest case of perpendicular $\vec{\lambda}$
and $\vec{B}$. In Fig. \ref{variob} we display the resulting phase diagram
for some parameters, showing the possible different shapes. There are two
gapped phases, the trivial (white region II) and the topological one (light
gray I), separated in general by two circular arcs defined by Eqs. 
(\ref{bound}). For simplicity we discuss the case $t, B > 0$. The topological character
is independent of the sign of the different parameters. If $B<2t$, the
region of possible values of $|\mu|$ inside the topological sector extends
from $2t-B$ to $2t+B$ for $\Delta \rightarrow 0$ and shrinks for increasing $%
\Delta$ until it reduces to the point $|\mu|=2t$ for $\Delta \rightarrow B$.
If $B=2t$, the semicircle touches the point $\mu=0$. For larger $B$, the
region $|\mu|< \sqrt{B^{2}-\Delta^{2}}-2t$ for $\Delta^2 < B^{2}-4t^2$ is
excluded from the topological region.

\begin{figure}[h!]
\begin{center}
\includegraphics[width=\columnwidth]{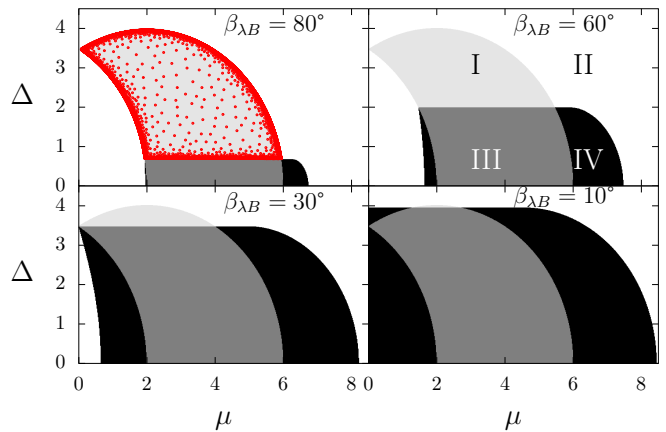}
\end{center}
\caption{(Color online) Phase diagram in the $\mu,\Delta$ plane for $t=1$, 
$\lambda=2$, $B=4$ and several values of the angle 
$\beta_{\lambda B}$ between $\vec{\lambda}$ and $\vec{B}$. 
Regions I and II as in Fig. \ref{variob}. Region III (IV)
in dark gray (black) corresponds to the gapless phase
with Berry phase $\pi$ (0).
The red points at the top left corresponds to numerical calculations
which detected localized states at the ends.}

\label{variobeta}
\end{figure}

While for perpendicular $\vec{\lambda}$ and $\vec{B}$, the gap vanishes only
at particular lines in the phase diagram (black lines in Fig. \ref{variob})
for which the topological transition takes place, for general angles $\beta
_{\lambda B}$ between both vectors, there is a finite region in the $\mu
,\Delta $ plane for which the gap vanishes, in particular for $|\Delta
|<\Delta _{c}$, where $\Delta _{c}$ is a critical value, independent of $\mu 
$, determined analytically below. Before presenting the analytical
calculation, we describe the general features of each phase in the phase
diagram, as shown in Fig. \ref{variobeta}. The gapped regions in the figure
are denoted by I and II. The remaining two regions are gapless. We separate them by the trivial (topological) character of the Berry phases of the second and third eigenstate, indicating the
corresponding regions with black (dark gray) color and roman number IV (III). 
In spite of the topological Berry phases of the latter gapless phase, 
MZMs in a finite chain are not expected to be protected against small
perturbations because of the absence of a gap. Therefore we describe this
phase as non topological. Furthermore we do not find numerically signatures
of localized end states in this phase.

We have also checked the boundaries of the
topological phase solving numerically finite chains and searching for
localized states at their ends and the presence of the finite gap. The
localized states are described in Sec. \ref{majo}. The presence of the
gap is defined by the condition that the determinant $D(k)$ of $H_{k}$ is
positive for each $k$. As it can be seen in Fig. \ref{variobeta} top left,
the results of both approaches agree.

\subsection{Analytical expressions for the boundaries of the topological
phase}

\label{anal}

For perpendicular $\vec{\lambda}$ and $\vec{B}$, the boundaries of the
topological phase are defined by Eqs. (\ref{bound}) and the conditions 
$|\vec{\lambda}|\neq 0$ and $\Delta \neq 0$. As the angle is changed 
from  90\textdegree, the gap reduces and a non-zero $|\Delta |$ is necessary to keep the gap
open (see Fig. \ref{variobeta}). For convenience, we discuss first the case $
\lambda _{z}=0$ (perpendicular $\vec{\lambda}$ and $\vec{B}$) and later
consider the general case $\vec{\lambda}=\lambda _{y}\mathbf{\hat{y}}
+\lambda _{z}\mathbf{\hat{z}}$ with $\lambda _{z}\neq 0$. For $\lambda _{z}=0
$, the determinant $D_{0}(k)$ of $H_{k}$ [see Eqs. (\ref{hk3}) or (\ref{hkp}
)]

\begin{eqnarray}
D_{0}(k) &=&C^{2}+4\Delta ^{2}y^{2},  \notag \\
C &=&a^{2}+\Delta ^{2}-B^{2}-y^{2},  \label{d0k}
\end{eqnarray}%
is positive semidefinite. It can vanish only for $y=0$ implying either $k=0$
or $k=\pi $. For $k=0$ ($k=\pi )$, $C=0$ implies $|\mu +2t|=r$ ($|\mu -2t|=r$
). Comparing with Eqs. (\ref{bound}), one realizes that the gap vanishes in
general only at one wave vector and only at the transition between
topological and non-topological gapped phases, as expected. The exception is
the case $|2t|=r$ and $\mu =0$, for which the gap vanishes at both wave
vectors.

In the general case with $z=2\lambda _{z}\sin (k)$ non zero, the determinant
of $H_{k}$ is [see Eq. (\ref{hkp})]

\begin{equation}
D(k)=D_{0}+2z^{2}(\Delta ^{2}+y^{2}-a^{2}-B^{2})+z^{4}  \label{dk}
\end{equation}
We can consider $D(k)$ as a function of $x=\cos (k)$. For large enough $
|\lambda _{z}|$, it turns out that, at the wave vector $k=0$, and parameters
for which $C=y=z=0$ [implying $D(0)=0$], $dD(x)/dx>0$ and as a consequence
for small positive $k$ ($x<1$) the determinant becomes negative signaling
the instability of the gapped phase. For $\lambda _{z}=0$, as in the previous
case the derivative is negative, but $x$ cannot be increased beyond 1, so
that $D(k)\geq 0$. A similar reasoning with the corresponding changes in the
sign can be followed for $k=\pi $. An explicit calculation of the derivative
using the conditions $C=\sin (k)=0$ gives 
\begin{equation}
\frac{dD}{dx}=32[B^{2}\lambda _{z}^{2}-\Delta ^{2}(\lambda _{z}^{2}+\lambda
_{y}^{2})]x.  \label{dddx}
\end{equation}
This implies that to have a gap one needs that $|\Delta |>\Delta _{c}$ where 
\begin{equation}
\Delta _{c}^{2}=B^{2}\frac{\lambda _{z}^{2}}{\lambda _{z}^{2}+\lambda
_{y}^{2}}=B^{2}\cos ^{2}(\beta _{\lambda B}).  \label{deltac}
\end{equation}
This condition has been found before for a model similar to ours in the
continuum with quadratic dispersion.\cite{rex}

After some algebra, the determinant in the general case can be written in
the form 
\begin{equation}
D=(C-z^{2})^{2}+16(\lambda _{z}^{2}+\lambda _{y}^{2})(\Delta ^{2}-\Delta
_{c}^{2})(1-x^{2}),  \label{d2}
\end{equation}
which is again positive semidefinite for $|\Delta |>\Delta _{c}$ and
positive definite for $0\neq k\neq \pi $, indicating a gapped phase. Since $
x=1$ implies $y=z=1$, the remaining boundaries of the topological phase
remain the same as for perpendicular $\vec{\lambda}$ and $\vec{B}.$ For $
|\Delta|=\Delta _{c}$ (as in Fig. \ref{eigenmu}), the values of $k$ for
which the determinant vanishes are given by the solutions with $|x|\leq 1$
of the following quadratic equation

\begin{eqnarray}
0 &=&4(t^{2}+\lambda ^{2})x^{2}+4t\mu x  \notag \\
&&+\mu ^{2}+\Delta _{c}^{2}-B^{2}-4\lambda ^{2},  \label{xc}
\end{eqnarray}
where $\lambda =|\vec{\lambda}|$.

\subsection{Transition from the topological phase to the gapless phases}

\label{topogap}

\begin{figure}[h!]
\begin{center}
\includegraphics[width=\columnwidth]{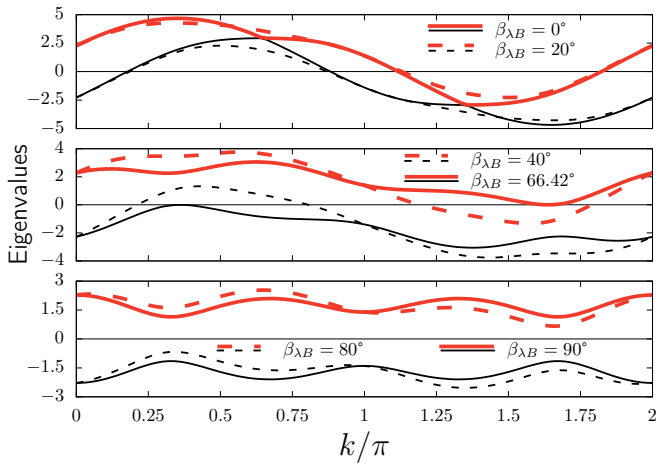}
\end{center}
\caption{(Color online) Second (black thin lines) and third (red thick lines) eigenvalues of $H_k$ as a function of wave vector
for $t=1$, $\lambda=\Delta=2$, $B=\mu=5$, and several values
of the angle $\beta_{\lambda B}$ between 
$\vec{\lambda}$ and $\vec{B}$.}
\label{eigenbeta}
\end{figure}

To gain insight into the transition from the topological phase to the
gapless phases, we represent in Fig. \ref{eigenbeta} the second and third
eigenvalues of $H_{k}$ [$E_{2}(k)$ and $E_{3}(k)$, respectively] for
different values $\beta _{\lambda B}$ of the angle between $\vec{\lambda}$
and $\vec{B}$. The parameters are such that, for $\vec{\lambda}\cdot \vec{B}=0
$, the system is in the topological phase with a finite gap. As the angle is
changed (in either direction) the gap between the second and third
eigenvalue decreases until at a certain critical angle [given by the
solution of Eq. (\ref{xc})] $E_{2}(k_{c})=E_{3}(-k_{c})=0$ at one particular
wave vector $k_{c}$ ($0.3613\pi$ in the figure), denoting the onset of the
gapless phase. Further turning $\vec{\lambda}$ and $\vec{B}$ to the parallel
(or antiparallel) direction, both eigenvalues vanish at two different wave
vectors.

\begin{figure}[h!]
\begin{center}
\includegraphics[width=\columnwidth]{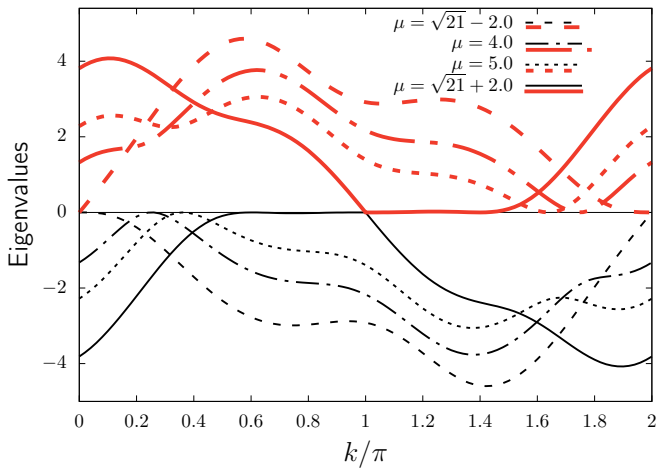}
\end{center}
\caption{(Color online) Same as Fig. \ref{eigenbeta} 
for $t=1$, $\lambda=\Delta=2$, $B=5$, 
$\beta_{\lambda B}=66.42$\textdegree \ and several values of $\mu$.}
\label{eigenmu}
\end{figure}

If keeping the other parameters fixed, the chemical potential $\mu $ is
changed towards one border $\mu _{c}$ of the topological phase for $\vec{%
\lambda}\cdot \vec{B}=0$ [given by Eq. (\ref{bound})]; the critical wave
vector $k_{c}$ is displaced either to $k_{c}=0$ or to $k_{c}=\pi $ depending
on the border. This is illustrated in Fig. \ref{eigenmu}. At the
corresponding border $\mu =\mu _{c}$,  one has $E_{2}(k_{c})=E_{3}(k_{c})=0$, indicating a crossing of the levels which is also accompanied by a change
in the Berry phases of the corresponding eigenvectors. Further displacing $%
\mu $ the system enters the non topological gapped phase. Therefore, the
point $\mu =\mu _{c}$,  $\Delta =\Delta _{c}$ is at the border of the
topological phase, the nontrivial gapless phase with Berry phase $\pi$, and the non-topological gapped phase. In fact also the trivial gapless phase reaches
this tetracritical point in the phase diagram (see Fig. \ref{variobeta}).

\subsection{Majorana modes}

\label{majo}

\begin{figure}[h!]
\begin{center}
\includegraphics[width=\columnwidth]{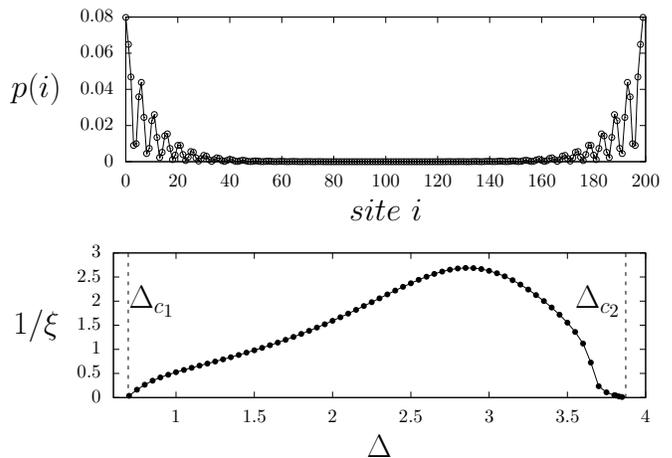}
\end{center}
	\caption{Top: probability of finding a fermion at each site of a chain  for the eigenstate of lowest positive energy for $t=1$, $\lambda=2$, $B=4$, $\beta_{\lambda B}=80$\textdegree, $\mu=3$, and $\Delta=0.75$. Bottom: inverse of the localization length as a function of $\Delta$. The transition between phases I and III is at $\Delta_{c_1}=0.694592711$ and the transition between phases I and II is at $\Delta_{c_2}=3.872983346$.}
\label{pis}
\end{figure}

The topological phase is characterized 
by the presence of Majorana modes zero modes at the ends of an infinite chain.
For a finite chain, the modes at both ends mix, giving rise to a 
fermion $\Gamma$ and its Hermitian conjugate with energies $\pm E$ which 
decay exponentially with the length $L$ of the chain. We have obtained $\Gamma$ numerically in chains of $L \sim 200$ sites. The probability $p(i)$ of finding
a fermion at site $i$ (adding both spins and creation and annihilation) is shown
in Fig. \ref{pis}. The main feature of the top figure is  a decay of $p(i)$
as the distance from any of the ends increases. We have chosen a case with a rather slow decay to facilitate visualization. In addition to this decay, some oscillations are visible with a short period.

In order to quantify the decay length of the localization of the end modes, we have fit the probability with an exponentially decaying function 
$p(i) \sim A$exp$(-i/\xi)$ at the left end. At the bottom of 
Fig. \ref{pis} we show the dependence
of $\xi$ inside the topological phase \textrm{I} as one of the parameters is varied.
As expected, $\xi$ diverges at the boundary with the non topological gapped phase \textrm{II}, which has a different $\mathbb{Z}_{2}$ topological invariant 
(at $\Delta_{c_2}=3.872983346$ in the figure). We also find that  
$\xi$ diverges at the boundary with the gapless phase \textrm{III} 
(at $\Delta_{c_1}=0.694592711$ in the figure), a phase with the same topological invariant but gapless. These facts allow us to obtain numerically
the transitions from the localization of the end states 
(see top left panel of Fig. \ref{variobeta}).

\section{Summary and discussion}

\label{sum}

Using numerical and analytical methods, 
we calculate the phase diagram of a widely used model for topological superconducting
wires, the essential ingredients of which are local s-wave pairing $\Delta$, spin-orbit coupling $\vec{\lambda}$ and magnetic field $\vec{B}$. 
We determine the boundary of the gapped topological phase analytically. 
This phase contains robust Majorana zero modes at both ends that are of great interest.
We expect that
this result will be relevant for future studies in the field.

The optimal situation for topological superconductivity is when $\vec{B}$ is perpendicular
to $\vec{\lambda}$. In this case, both the topological and non-topological phases are gapped. If instead $\vec{B}$ has a component in the direction of $\vec{\lambda}$, a gapless
superconducting phase appears for certain parameters. This phase can also be separated in two 
phases differing in a $\mathbb{Z}_{2}$ topological invariant. However, due to the absence of a gap,
we do not find Majorana zero-modes at the ends of the phase with 
nontrivial $\mathbb{Z}_{2}$, in contrast to those present in the gapped topological phase.

Tilting the magnetic field to enter the gapless phase might be used as a trick to relax
the system to the ground state in some measurements, like Josephson current.
In the gapped topological phase, in the absence of 
low-frequency phonons or other excitations, the physics is dominated by a few 
bound states inside the gap, completely isolated from the continuum, and the current would oscillate, without reaching a steady state.\cite{chung} One way to avoid this problem would be to use a magnetic field so that the system is in the gapless phase, with low-energy
excitations available for thermalization,
and then rotate adiabatically the field to the desired value so that the system remains in the ground state.

\section*{Acknowledgments}

We thank L. Arrachea for helpful discussions. We are sponsored by PIP
112-201501-00506 of CONICET, PICT-2017-2726, PICT-2018-04536 and
PICT-Raices-2018.


\begin{thebibliography}{99}
\bibitem{sato} M. Sato and Y. Ando, Topological superconductors: a review,
Rep. Prog. Phys. \textbf{80}, 076501 (2017).

\bibitem{kitaev} A. Kitaev, Fault-tolerant quantum computation by anyons,
Ann. Phys. (N.Y.) \textbf{303}, 2 (2003).

\bibitem{nayak} C. Nayak, S. H.Simon, A. Stern, M. Freedman, and S. Das
Sarma, Non-Abelian anyons and topological quantum computation, Rev. Mod.
Phys. \textbf{80}, 1083 (2008).

\bibitem{alicea} J. Alicea, New directions in the pursuit of Majorana
fermions in solid state systems, Rep. Prog. Phys. \textbf{75}, 076501,
(2012).

\bibitem{lobos} X-J. Liu and A. M. Lobos, Manipulating Majorana fermions in
quantum nanowires with broken inversion symmetry, Phys. Rev. B \textbf{87},
060504(R), (2013).

\bibitem{lutchyn2010} R. M. Lutchyn, J. Sau, and S. Das Sarma, Majorana
Fermions and a Topological Phase Transition in Semiconductor-Superconductor
Heterostructures, Phys. Rev. Lett. \textbf{105} 077001 (2010).

\bibitem{oreg2010} Y. Oreg, G. Refael, and F. von Oppen, Helical Liquids and
Majorana Bound States in Quantum Wires, Phys. Rev. Lett. \textbf{105} 177002
(2010).

\bibitem{wires-exp1} V. Mourik, K. Zuo, S. M. Frolov, S. R. Plissard, E. P.
a. M. Bakkers, and L. P. Kouwenhoven, Signatures of Majorana fermions in in
hybrid superconductor-semiconductor nanowire devices, Science \textbf{336},
1003 (2012).

\bibitem{wires-exp2} A. Das, Y. Ronen, Y. Most, Y. Oreg, M. Heiblum, and H.
Shtrikman, Zero-bias peaks and splitting in an Al-InAs nanowire topological
superconductor as a signature of Majorana fermions, Nat. Phys. \textbf{8},
887 (2012).

\bibitem{wires-exp3} S. M. Albrecht, A. P. Higginbotham, M. Madsen, F.
Kuemmeth, T. S. Jespersen, J. Nyg, P. Krogstrup, and C. M. Marcus,
Exponential protection of zero modes in Majorana islands, Nature \textbf{531}%
, 206 (2016).

\bibitem{wires-exp4} M. Deng, S. Vaitiekenas, E. Hansen, J. Danon, M.
Leijnse, K. Flensberg, J. Nyg\aa rd, P. Krogstrup, and C. Marcus, Majorana
bound state in a coupled quantum-dot hybrid-nanowire system, Science \textbf{%
354}, 1557 (2016).

\bibitem{review-tritops} A. Haim and Y.Oreg, Time-reversal-invariant
topological superconductivity in one and two dimension, Phys. Rep. \textbf{%
825}, 1 (2019).

\bibitem{volpez} Y. Volpez, D. Loss, and J. Klinovaja, Time-reversal
invariant topological superconductivity in planar Josephson bijunction,
Phys. Rev. Research \textbf{2}, 023415 (2020).

\bibitem{zazu} A. Zazunov, R. Egger, and A. Levy Yeyati, Low-energy theory
of transport in Majorana wire junctions, Phys. Rev. B \textbf{94}, 014502
(2016).

\bibitem{pientka} F. Pientka, A. Keselman, E. Berg, A. Yacoby, A. Stern, and
B. I. Halperin, Topological Superconductivity in a Planar Josephson
Junction, Phys. Rev. X \textbf{7}, 021032 (2017).

\bibitem{hell} M. Hell, M. Leijnse, and K. Flensberg, Two-Dimensional
Platform for Networks of Majorana Bound States, Phys. Rev. Lett. \textbf{118}%
, 107701 (2017).

\bibitem{ren} H. Ren, F. Pientka, S. Hart, A. Pierce, M. Kosowsky, L.
Lunczer, R. Schlereth, B. Scharf, E. M. Hankiewicz, L. W. Molenkamp, B. I.
Halperin, and A. Yacoby, Topological superconductivity in a phase-controlled
Josephson junction, Nature \textbf{569}, 93 (2019).

\bibitem{fornie} A. Fornieri, A. M. Whiticar, F. Setiawan, E. P. Martin, A.
C. C. Drachmann, A. Keselman, S. Gronin, C. Thomas, T. Wang, R. Kallaher, G.
C. Gardner, E. Berg, M. J. Manfra, A. Stern, C. M. Marcus, and F. Nichele,
Evidence of topological superconductivity in planar Josephson junctions,
Nature 569, \textbf{89} (2019).

\bibitem{cata} L. Arrachea, A. Camjayi, A. A. Aligia, and L. Gru\~neiro,
Catalog of Andreev spectra and Josephson effects in structures with
time-reversal-invariant topological superconductor wires, Phys. Rev. B 
\textbf{99}, 085431 (2019).

\bibitem{tomo} A. A. Aligia, D. P\'{e}rez Daroca, and L. Arrachea,
Tomography of Zero-Energy End Modes in Topological Superconducting Wires,
Phys. Rev. Lett. \textbf{125}, 256801 (2020).

\bibitem{bondy} N. Bondyopadhaya and D. Roy, Dynamics of hybrid junctions of
Majorana wires, Phys. Rev. B \textbf{99}, 214514 (2019).

\bibitem{tewari} S. Tewari and J. D. Sau, Topological invariants for
spin-orbit coupled superconductor nanowires, Phys. Rev. Lett. \textbf{109},
150408 (2012).

\bibitem{zak} J. Zak, Berry's phase for energy bands in solids, Phys. Rev.
Lett. \textbf{62}, 2747 (1989).

\bibitem{king} R. D. King-Smith and David Vanderbilt, Theory of polarization
of crystalline solids, Phys. Rev. B \textbf{47}, 1651 (1993).

\bibitem{resta} R. Resta and S. Sorella, Many-body effects on polarization
and dynamical charges in a partly covalent polar insulator, Phys. Rev. Lett. 
\textbf{74}, 4738 (1995).

\bibitem{ortiz} G. Ortiz, P. Ordej\'{o}n, R. M. Martin, and G. Chiappe,
Quantum phase transitions involving a change in polarization, Phys.Rev. B 
\textbf{54}, 13515 (1996).

\bibitem{bf1} A. A. Aligia, K. Hallberg, C. D. Batista and G. Ortiz, Phase
diagrams from topological transitions: The Hubbard chain with correlated
hopping, Phys. Rev. B \textbf{61}, 7883 (2000).

\bibitem{bf2} A. A. Aligia, K. Hallberg, B. Normand, and A. P. Kampf,
Detection of Topological Transitions by Transport Through Molecules and
Nanodevices, Phys. Rev. Lett. \textbf{93}, 076801 (2004).

\bibitem{hatsu} Y. Hatsugai, Quantized Berry Phases as a Local Order
Parameter of a Quantum Liquid. Journal of the Physical Society of Japan 
\textbf{75}, 123601 (2006).

\bibitem{bf3} A. A. Aligia, A. Anfossi, L. Arrachea, C. Degli Esposti
Boschi, A. O. Dobry, C. Gazza, A. Montorsi, F. Ortolani, and M. E. Torio,
Incommmensurability and Unconventional Superconductor to Insulator
Transition in the Hubbard Model with BondCharge Interaction, Phys. Rev.
Lett. \textbf{99}, 206401 (2007).

\bibitem{Schn} A. P. Schnyder, S. Ryu, A. Furusaki, and A. W. W. Ludwig,
Classification of topological insulators and superconductors in three spatial dimensions,
Phys. Rev. B \textbf{78}, 195125 (2008). 

\bibitem{ryu} S. Ryu, A. P. Schnyder, A. Furusaki, and A. W. W. Ludwig,
Topological insulators and superconductors: tenfold way and dimensional
hierarchy, New Journal of Physics \textbf{12}, 065010 (2010).

\bibitem{deng} S. Deng, G. Ortiz, and L. Viola, Multiband $s$-wave
topological superconductors: Role of dimensionality and magnetic field
response. Phys. Rev. B \textbf{87}, 205414 (2013).

\bibitem{budich} J. C. Budich and E. Ardonne, Equivalent topological
invariants for one-dimensional Majorana wires in symmetry class D, Phys.
Rev. B \textbf{88}, 075419 (2013).

\bibitem{rex} S. Rex and A. Sudbo, Tilting of the magnetic field in Majorana
nanowires: Critical angle and zero-energy differential conductance, Phys.
Rev. B \textbf{90}, 115429 (2014).

\bibitem{chung} S. B. Chung, J. Horowitz, and X-L. Qi, Time-reversal anomaly
and Josephson effect in time-reversal-invariant topological superconductors,
Phys. Rev. B \textbf{88}, 214514 (2013).
\end{thebibliography}
\end{document}